\documentclass[11pt,twoside]{article}
\usepackage{amssymb}
\usepackage{graphicx}
\usepackage{graphicx,type1cm,eso-pic,color}
\usepackage{amssymb}
\usepackage{amssymb,amsmath}
\usepackage{graphicx,epsfig}
\usepackage{enumerate}
\usepackage{color}
\usepackage {multicol}
\usepackage{lineno}
\numberwithin{equation}{section}
\newtheorem{theorem}{Theorem}
\newtheorem{lemma}{Lemma}

\newtheorem{remark}{Remark}

\newtheorem{example}{Example}

\makeatletter
\AddToShipoutPicture{
	\setlength{\@tempdimb}{.5\paperwidth}
	\setlength{\@tempdimc}{.5\paperheight}
	\setlength{\unitlength}{1pt}
	\put(\strip@pt\@tempdimb,\strip@pt\@tempdimc)
}
\makeatother

\usepackage{scalerel,stackengine}
\stackMath
\newcommand\reallywidehat[1]{%
	\savestack{\tmpbox}{\stretchto{%
			\scaleto{%
				\scalerel*[\widthof{\ensuremath{#1}}]{\kern-.6pt\bigwedge\kern-.6pt}%
				{\rule[-\textheight/2]{1ex}{\textheight}}
			}{\textheight}%
		}{0.5ex}}%
	\stackon[1pt]{#1}{\tmpbox}%
}
\begin{document}
	\setcounter{page}{1}

	\thispagestyle{empty}
	\markboth{}{}

	\pagestyle{myheadings}
	\markboth{ S.K.Chaudhary and N.Gupta }{ S.K.Chaudhary and N.Gupta }
	
	\date{}
	
	
	\noindent  
	
	\vspace{.1in}
	
	{\baselineskip 20truept
		
		\begin{center}
			{\Large {\bf Testing exponentiality using extropy of upper record values}} \footnote{\noindent	{\bf *} Corresponding author E-mail: skchaudhary1994@kgpian.iitkgp.ac.in\\
				{\bf ** }  E-mail: nitin.gupta@maths.iitkgp.ac.in }\\
			
	\end{center}}

	\vspace{.1in}
	
	\begin{center}
		{\large {\bf Santosh Kumar Chaudhary* and Nitin Gupta**}}\\
		{\large {\it Department of Mathematics, Indian Institute of Technology Kharagpur, West Bengal 721302, India. }}
		\\
	\end{center}

	\vspace{.1in}
	\baselineskip 12truept
	\begin{center}
		{\bf \large Abstract}\\
	\end{center}
	We are giving one characterization result of exponential distribution using extropy of $n$th upper $k$-record value. We introduce test statistics based on the proposed characterization result that will be used to test exponentiality. The critical value and power of the test have been calculated using monte Carlo simulation. The test is applied to seven real-life data sets to verify its applicability in practice. 

	\section{Introduction}
	Let $X_1, X_2, ..., X_N$ be a random sample of size $N$ from a population with unknown probability density function(pdf) $f$ and cumulative distribution function (cdf) $F$. We consider $X$ as a non-negative random variable. A probability distribution is said to be exponential with parameter $\lambda$ if 
	\begin{align}\label{exppdfdef}
		f(x)=\lambda e^{-\lambda x},\ x>0, \ \lambda >0.
	\end{align}
	Here, we are interested in testing whether a distribution is exponential. That is,	
	$H_0:$ X has exponential($\lambda$) distribution  against 
	$H_1:$ X does not have exponential($\lambda$) distribution .

	\noindent Extropy of X is defined by Lad et al. (2015) is 
	\begin{align}\label{extropydef}
		J(X)=-\frac{1}{2} \int_{0}^{\infty} f^2(x)dx.
	\end{align}
	The cumulative residual extropy (CRE) of X is defined by  Jahanshahi et al. (2020) is
	\begin{align}\label{credef}
		\xi J(X)=-\frac{1}{2} \int_{0}^{\infty} \bar{F}^2_X(x)dx.
	\end{align}
	The cumulative past extropy (CPE) of X is defined as 
	\begin{align}\label{cpedef}
		\bar{\xi} J(X)=-\frac{1}{2} \int_{0}^{\infty} F^2_X(x)dx.
	\end{align}
	Let $X_1, X_2, \dots, X_N, \dots$ be a sequence of independent and identically distributed (iid) random variables from an absolutely continuous cdf F and pdf f. Let $X_{r:N}$ be $r$th order statistics for $0\leq r\leq N$ which is $r$th smallest in the sequence $X_1, X_2, \dots, X_N.$ Sequence of upper record time U(k) is defined as $U(1)=1, U(k+1)=min\{j:j>U(k);\ X_j>U^X_k\},\ k=1,2,3,\dots$ and the kth upper record value $U_k$ is defined as $U_k=X_{U(k)}.$ The pdf of $k$th upper record value $U_k$ is given as:
	\begin{align}
		f_{U_k}(x)=\frac{1}{\Gamma(k)}(-\log \bar{F}(x))^{k-1}  f(x),
	\end{align}
	where $\bar{F}(x)=1-F(x)$ is a survival function of X and $\Gamma(k)=(k-1)!$ is a complete gamma function. An analogous definition can be given to lower $k$-record value (see Arnold et al.(1998))\cite{arnoldbala98}. The pdf of kth lower record value $L_k$ is
	\begin{align}\label{pdflkx}
		f_{L_k}(x)=\frac{1}{\Gamma(k)}(-\log{F}(x))^{k-1}  f(x). 
	\end{align}
	The pdf of the $n$th upper $k$-record value $U_{n,k}$ and the $n$th lower $k$-record value $L_{n,k}$ respectively are given by (see Arnold et al.(1998 \cite{arnoldbala98}, 2008 \cite{arnoldbalanaga08}) ) 
	
	\begin{align}
		&f_{U_{n,k}}(x)=\frac{k^n}{\Gamma(n)}(-\log \bar{F}(x))^{n-1} (\bar{F}(x))^{k-1}f_X(x),\label{pdfurv}\\
		\text{and} \ \ \ &f_{L_{n,k}}(x)=\frac{k^n}{\Gamma(n)}(-\log F_X(x))^{n-1} (F_X(x))^{k-1}f_X(x).\label{pdflrv}
	\end{align}
	
	\noindent The cdf of $U_{n,k}$ and  $L_{n,k}$, respectively, are  
	\begin{align}
		F_{U_{n,k}}(x)&=  1-\bar{F}^k_X (x)   \sum_{i=0}^{n-1} \frac{(-k \log\bar{F}_X (x) )^i}{i!},\label{cdfurv}\\
		\text{and} \ \ \ 	F_{L_{n,k}}(x)&=  {F}^k_X (x)   \sum_{i=0}^{n-1} \frac{(-k \log F_X (x) )^i}{i!}\label{cdflrv}.
	\end{align}
	Xiong et al. (2020)\cite{xiongetalexp20} proposed test statistics based on the characterization of exponential distribution using extropy of upper $k$-record value. Jose and Sathar (2022)\cite{josesathaeexp22} proposed test statistics based on the characterization of exponential distribution using extropy of $n$th lower $k$-record value. In this paper, we present some more characterization of exponential distribution and introduce test statistics based on that characterization. Section 2 of this paper discusses some examples and theorems for the exponential distribution.
	In Section 3, we obtain test statistics for testing exponentiality.
	
	\section{Characterization of exponential distribution }
	Let us discuss some examples before we proceed to the main result of this section.
	\begin{example}\label{example1}
		When $X$ is exponential random variable with parameter $\lambda>0$,  and cdf  $F_X(x)=1-e^{-\lambda x},\ x>0,$ then extropy of $X,$ 	 $$J(X)=-\frac{1}{2} \int_{0}^{\infty} f^2(x)dx= -\frac{1}{2} \int_{0}^{\infty} \lambda^2 e^{-2\lambda x}dx =-\frac{\lambda}{4}.$$
	\end{example}
	
	\begin{example}\label{example2}(Xiong et al. (2020)\cite{xiongetalexp20})
		When $X$ is exponential random variable with parameter $\lambda>0$,  and cdf  $F_X(x)=1-e^{-\lambda x},\ x>0,$ then extropy of $U^X_k$ is 	 $$J(U^X_k)=\frac{-\lambda \Gamma(2k-1)}{2^{2k}\Gamma^2(k)} =\frac{\Gamma(2k-1)}{2^{2k-2}\Gamma^2(k)} J(X) $$
	\end{example}
	Theorem 2.1 of Xiong et al. (2020)\cite{xiongetalexp20} tells that a non-negative random variable $X$ has an exponential distribution with a rate
	parameter $\lambda>0$ if and only if $$J(U^X_k)= \frac{\Gamma(2k-1)} {2^{2k-2} \Gamma^2(k)} J(X), \ k=1,2, \dots.$$ 
	Also, Xiong et al. (2020)\cite{xiongetalexp20} proposed a goodness of fit test for exponentiality based on the above characteristics and analysed the performance of the proposed test statistics.

	\begin{example}\label{example3} (Jose and Sathar (2022)\cite{josesathaeexp22})
		When $X$ is exponential random variable with parameter $\lambda>0$,  and cdf  $F_X(x)=1-e^{-\lambda x},\ x>0,$ then extropy of $J(L_{n,k})$ is
		\begin{align*}
			J(L_{n,k})=&\frac{-\lambda k\Gamma(2n-1)}{2^{2n}\Gamma^2(n)} \left(\left(\frac{2k}{2k-1}\right)^{2n-1}-1\right) \\
			=&\frac{kJ(X)\Gamma(2n-1)}{2^{2n-2}\Gamma^2(n)} \left(\left(\frac{2k}{2k-1}\right)^{2n-1}-1\right).
		\end{align*}
	\end{example}
	Theorem 2.1 of Jose and Sathar (2022)\cite{josesathaeexp22} proved that a non-negative random variable $X$ has an exponential distribution with a rate parameter $\lambda>0$ if and only if $$J(L_{n,k})=\frac{kJ(X)\Gamma(2n-1)}{2^{2n-2}\Gamma^2(n)} \left(\left(\frac{2k}{2k-1}\right)^{2n-1}-1\right).$$\\
	Also, Jose and Sathar (2022)\cite{josesathaeexp22} proposed a test for exponentiality based on the above characteristics and analysed the performance of the proposed test statistics. \\
	The following example is obtained as a particular case of Example \ref{example3}.
	\begin{example}\label{example4} (Jose and Sathar (2022)\cite{josesathaeexp22})
		When $X$ is exponential random variable with parameter $\lambda>0$,  and cdf  $F_X(x)=1-e^{-\lambda x},\ x>0,$ then extropy of $J(L^X_k)$ is 
		\begin{align*}
			J(L^X_k)=& \frac{-\lambda \Gamma(2k-1)(2^{2k-1}-1)}{2^{2k} \Gamma^2(k)} \\
			=& \frac{\Gamma(2k-1)(2^{2k-1}-1)}{2^{2k-2} \Gamma^2(k)} J(X)
		\end{align*}
	\end{example}
	
	\begin{example}\label{example5}
		When $X$ is exponential random variable with parameter $\lambda>0$,  and cdf  $F_X(x)=1-e^{-\lambda x},\ x>0.$  The extropy of $U_{n,k}$ is 
		\begin{align*}
			J(U_{n,k}) &=\frac{-1}{2} \int_{0}^{\infty} f^2_{U_{n,k}}(x)dx \\
			&= \frac{-1}{2} \int_{0}^{\infty} \left(\frac{k^n}{\Gamma(n)}(-\log \bar{F}(x))^{n-1} (\bar{F}(x))^{k-1}f_X(x)\right)^2dx \\
			&= \frac{-1}{2} \frac{k^{2n}}{\Gamma^2(n)} \int_{0}^{\infty} \left((-\log \bar{F}(x))^{2n-2} (\bar{F}(x))^{2k-2}f^2_X(x)\right)dx.
		\end{align*}
		Since X has exponential distribution. Therefore putting $\bar{F}(x)= e^{-\lambda x}$, $f(x)=\lambda e^{-\lambda x}$  and $J(X)= \frac{-\lambda}{4}$, we have
		\begin{align*}
			J(U_{n,k})=& \frac{-1}{2}\frac{k^{2n}}{\Gamma^2(n)} \int_{0}^{\infty} (\lambda x)^{2n-2} (e^{-\lambda x})^{2k-2} \lambda^2 e^{-2\lambda x} dx \\
			=& \frac{-\lambda k\Gamma(2n-1)}{2^{2n}\Gamma^2(n)}\\
			=& \frac{k\Gamma(2n-1)}{2^{2n-2}\Gamma^2(n)} J(X).
		\end{align*} 
	\end{example}

	\begin{example}\label{example6}
		When $X$ is exponential random variable with parameter $\lambda>0$,  and cdf  $F_X(x)=1-e^{-\lambda x},\ x>0.$ The CRE of $n$th upper $k$-record value $U_{n,k}$ is 
		\begin{align*}
			\xi J(U_{n,k})=&\frac{-1}{2} \int_{0}^{\infty} \bar F_{U_{n,k}}^2(x)dx\\
			=&\frac{-1}{2} \int_{0}^{\infty} \left( \bar F^k(x)\sum_{i=0}^{n-1}\frac{\left(-k \log \bar F(x)\right)^i}{i!}\right)^2dx
		\end{align*}
		Since X has exponential distribution. Therefore putting $\bar{F}(x)= e^{-\lambda x}$ and $J(X)= \frac{-\lambda}{4}$, we get
		\begin{align*}
			\xi J(U_{n,k})&=\frac{-1}{2} \int_{0}^{\infty} \left( (e^{-\lambda x})^k \sum_{i=0}^{n-1}\frac{\left(-k \log e^{-\lambda x}\right)^i}{i!}\right)^2dx\\
			&=\frac{-1}{2} \sum_{i=0}^{n-1}\sum_{j=0}^{n-1} \frac{(k \lambda)^{i+j}}{i!j!} \int_{0}^{\infty} e^{-2k \lambda x}x^{i+j} dx\\
			&=\frac{-1}{4\lambda k}\sum_{i=0}^{n-1}\sum_{j=0}^{n-1}\frac{(i+j)!}{i!j!}\left(\frac{1}{2}\right)^{i+j}\\
			&= \frac{1}{16k J(X) } \sum_{j=0}^{n-1} \sum_{i=0}^{n-1}\frac{(i+j)!}{i!j!}\left(\frac{1}{2}\right)^{i+j}.
		\end{align*}
	\end{example}
	
	We state the following lemma from  Goffman and Pedrick (2017) \cite{goffped17} which helps us to establish a characterization result for the exponential distribution.
	
	\begin{lemma} \label{lemma1polynomial} (Goffman and Pedrick (2017) \cite{goffped17})
		A complete orthonormal system for the space $L_2(0, +\infty)$ is given by the sequence of Laguerre function $\phi_n(x)=e^{\frac{-x}{2}} \frac{L_n(x)}{n!}, \ n\geq0,$ where $L_n(x)$ is the Laguerre polynomial defined as the sum of coefficients of $e^{-x}$ in the $n$th derivative of $x^n e^{-x}$, that is $L_n(x)=e^x \frac{d^n}{dx^n}(x^n e^{-x})=\sum_{i=0}^{n} (-1)^i {n \choose i}n(n-1)\dots(i+1)x^i.$ The meaning of the completeness of Laguerre functions in $L_2(0, +\infty)$ is that if $f\in L_2(0, +\infty)$ and $\int_{0}^{+\infty}f(x)e^{-\frac{x}{2}}L_n(x)dx=0 $  for all $n \geq 0$, then $f$ is zero almost everywhere.
	\end{lemma}
	
	\begin{remark}
		Lemma \ref{lemma1polynomial} is also used to prove Theorem 3.7 in Qiu (2017) \cite{qiuexorderrecord17}, Theorem 2.1 in Xiong et al. (2020) \cite{xiongetalexp20} and Theorem 2.1 in Jose and Sathar (2022) \cite{josesathaeexp22}.
	\end{remark}
	
	\begin{theorem}\label{thmjunk}
		A non-negative random variable X has an exponential distribution with rate parameter $\lambda>0$  if and only if $$J(U_{n,k})=\frac{k\Gamma(2n-1)}{2^{2n-2}\Gamma^2(n)} J(X), \ k=1,2, \dots.$$
	\end{theorem}
	\textbf{Proof:} Proof of necessity follows from Example \ref{example5}. The proof of sufficiency is similar to the proof of Theorem 2.1 of Jose and Sathar (2022)\cite{josesathaeexp22} and we present here. Suppose X is a non negative random variable such that $$J(U_{n,k})=\frac{k\Gamma(2n-1)}{2^{2n-2}\Gamma^2(n)} J(X).$$
	Using definition of extropy given in (\ref{extropydef}) and pdf of $U_{n,k}$ given in (\ref{pdfurv}), we have, 
	$$\int_{0}^{\infty} \left(\log(\bar{F}(x))\right)^{2n-2} (\bar{F}(x))^{2k-2}f^2(x)dx=-\frac{2J(X)\Gamma(2n-1)}{2^{2n-2} k^{2n-1}}.$$
	Now, after substituting $\log(\bar{F}(x))= -u$, we get 
	\begin{align}\label{eqn2.1}
		\int_{0}^{\infty}u^{2n-2} (e^{-u})^{2k-1} f(\bar{F}^{-1}(e^{-u}))du= -\frac{4J(X)\Gamma(2n-1)}{(2k)^{2n-1}}.
	\end{align}
	Using the result $\frac{\Gamma(\alpha)}{k^{\alpha}}=\int_{0}^{\infty} e^{-ku} u^{\alpha-1}du$, expression (\ref{eqn2.1}) reduces to 
	\begin{align}	
		&\int_{0}^{\infty} e^{-2ku} u^{2n-2} \{e^u f(\bar{F}^{-1}(e^{-u}))+ 4J(X)\}du=0.
	\end{align}
	We can write the above equation as 
	\begin{align}	
		&\int_{0}^{\infty}u^{n-1} e^{-u(2k+\frac{1}{2})} \{e^{2u} f(\bar{F}^{-1}(e^{-u}))+ 4e^u J(X)\}e^{-\frac{u}{2}}L_n(u)=0,
	\end{align}
	where $L_n(u)$ is Lagurre polynomial defined in lemma \ref{lemma1polynomial}. Therefore, \\ $u^{n-1}e^{-u(2k+\frac{1}{2})} \{e^{2u}f(\bar{F}^{-1}(e^{-u}))+ 4e^u J(X)\} \in L_2(0,+\infty).$  Using the completeness property in lemma \ref{lemma1polynomial}, we get,
	\begin{align} \label{eqnuv2.2}
		e^{2u} f(\bar{F}^{-1}(e^{-u}))+ 4e^u J(X)=0 \nonumber\\
		\text{that is,\ }
		e^uf(\bar{F}^{-1}(e^{-u})) +4J(X)=0.
	\end{align}
	If we take $v=e^{-u}$ then expression \ref{eqnuv2.2} reduces to 
	\begin{align} 
		f(\bar{F}^{-1}(v)) +4vJ(X)=0.
	\end{align}
	Further, substituting, $t=\bar{F}^{-1}(v)$, and using $\bar{F}(0)=1$, we get 
	$$\bar{F}(t)=e^{4J(X)t}, \ t>0.$$
	Hence, random variable X has an exponential distribution with rate parameter $-4J(X).$

	\section{Test Statistics}
	We introduce a test statistics $\Delta_{n,k}$ for testing exponentiality by using Theorem \ref{thmjunk} as given below
	\begin{align}\label{teststatisticsnk}
		\Delta_{n,k}= J(U_{n,k})-\frac{k\Gamma(2n-1)}{2^{2n-2}\Gamma^2(n)} J(X)
	\end{align}
	From Theorem \ref{thmjunk}, $\Delta_{n,k}=0$ for all $k= 1,2,3, \dots\,$ if and only if  X is exponential. Therefore, $\Delta_{n,k}$ is suitable to consider test statistics in testing the exponentiality of X. For computation purposes, we choose n=2 and k=2 for further discussion. One may choose another value for n and k, but the procedure remains the same. Now, Test statistics is 
	\begin{align}\label{teststatistics22}
		\Delta_{2,2}= J(U_{2,2})-J(X)
	\end{align}
	An estimator of $\Delta_{2,2}$ is 
	\begin{align}\label{testestimtor22}
		\reallywidehat{\Delta_{2,2}}=\reallywidehat{J(U_{2,2})}- \reallywidehat{J(X)}
	\end{align}
	\begin{remark}
		The test statistics $E_k$ introduced by Xiong et al.(2020) \cite{xiongetalexp20} is a particular case of our proposed test statistics $\Delta_{n,k}.$
	\end{remark}
	\begin{remark}
		Test statistics proposed by Xiong et al.(2020) \cite{xiongetalexp20} is 		$$E_k=J(U^X_k)-\frac{\Gamma(2k-1)}{2^{2k-2}\Gamma^2(k)} J(X).$$ For simplicity of calculation, they choose k=2, We get  $E_2=J(U^X_2)-\frac{1}{2} J(X)$ but the estimator considered in Xiong et al.(2020) \cite{xiongetalexp20}  for testing is $\reallywidehat{E_2}=\reallywidehat{J(U^X_2)}-\reallywidehat{J(X)}.$
		The test statistics and estimator should be  $E_2=J(U^X_2)-\frac{1}{2} J(X)$ and $\reallywidehat{E_2}=\reallywidehat{J(U^X_2)}- \frac{1}{2} \reallywidehat{J(X)}$ respectively.
	\end{remark}
	The Extropy of $X$ can be expressed as
	\begin{align*}
		J(X)=-\frac{1}{2} \int_{0}^{\infty} f^2(x)dx=-\frac{1}{2} \int_{0}^{1} \left(\frac{d}{du} F^{-1}(u)\right)^{-1} du
	\end{align*}
	Qui and Jia (2018) \cite{qiujia2018} introduced the sample estimator of $J(X)$ by $\reallywidehat{J(X)}$ as:
	\begin{align*}
		\reallywidehat{J(X)}=\frac{-1}{2N} \sum_{i=1}^{N} \frac{c_i m/N}{X_{i+m:N}-X_{i-m:N}}
	\end{align*}
	where $X_{1:N}, X_{2:N}, X_{3:N},\dots,X_{N:N}$ are order statistics based on $X_1,X_2,X_3,\\ \dots,X_N,$ and 
	\begin{align*}
		c_i=
		\begin{cases}
			1+\frac{i-1}{m} &\text{if $ 1\leq i \leq m,$}\\
			2 &\text{if $ m+1\leq i \leq N-m,$}\\
			1+\frac{N-i}{m} &\text{if $ N-m+1\leq i \leq N.$}
		\end{cases}
	\end{align*}\label{cidefine}
	
	The window size m is positive integer smaller than $\frac{N}{2},$ and if $i-m<1$ then  $X_{i-m:N}=X_{1:N}$ and if $i+m>N$ then $X_{i+m:N}=X_{N:N}.$ 
	
	Extropy of $U_{n,k}$ can be expressed as
	\begin{align*}
		J(U_{n,k}) &=\frac{-1}{2} \int_{0}^{\infty} f^2_{U_{n,k}}(x)dx \\
		&= \frac{-k^{2n}}{2\Gamma^2(n)} \int_{0}^{\infty} \left((\log \bar{F}(x))^{2n-2} (\bar{F}(x))^{2k-2}f^2_X(x)\right)dx
	\end{align*}
	After Substituting $u=F(x)$, we get
	\begin{align*}
		J(U_{n,k}) =&  \frac{-k^{2n}}{2\Gamma^2(n)} \int_{0}^{1} \left((\log (1-u))^{2n-2} ((1-u))^{2k-2} \left(\frac{d}{du} F^{-1}(u)\right)^{-1} \right)du\\
		J(U_{2,2}) =& -8 \int_{0}^{1} \left((\log (1-u))^{2} ((1-u))^{2} \left(\frac{d}{du} F^{-1}(u)\right)^{-1} \right)du
	\end{align*}
	Vasicek (1976), \cite{vasicek} developed a concept to find an estimator, and in accordance with that, an estimator of $J(U_{2,2})$ will be produced by substituting the empirical distribution function $\reallywidehat{F}_N$ for the distribution function $F$ and using the difference operator in place of a differential operator. The derivative of $F^{-1} (u)$ with respect to $u$, that is, $\frac{dF^{-1} (u)}{du}$ will be estimated as 
	\begin{align*}
		\frac{X_{i+m:N}-X_{i-m:N}}{\reallywidehat{F}_N(X_{i+m:N})-\reallywidehat{F}_N(X_{i-m:N}}=\frac{X_{i+m:N}-X_{i-m:N}}{\frac{i+m}{N}-\frac{i-m}{N}}=\frac{X_{i+m:N}-X_{i-m:N}}{2m/N}.  
	\end{align*}
	Analogous to Vasicek (1976) \cite{vasicek}, Park (1999) \cite{park}, Xiong et al.(2020) \cite{xiongetalexp20}, Jose and Sathar (2022)  \cite{josesathaeexp22},	we write estimator of $J(U_{n,k})$ and $J(U_{2,2})$, respectively, as follows,
	\begin{align*}
		\reallywidehat{J(U_{n,k})} =&  \frac{-k^{2n}}{2N\Gamma^2(n)} \sum_{i=1}^{N} \left((\log (1-\frac{i}{N+1}))^{2n-2} ((1-\frac{i}{N+1}))^{2k-2} \left(\frac{2m/N}{X_{i+m:N}-X_{i-m:N}}\right)\right)\\
		\reallywidehat{J(U_{2,2})} =&\frac{-8}{N}\sum_{i=1}^{N}\left((\log(1-\frac{i}{N+1}))^{2}(1-\frac{i}{N+1})^{2}\left(\frac{2m/N}{X_{i+m:N}-X_{i-m:N}}\right)\right)
	\end{align*}
	A reasonable estimator of $\Delta_{2,2}$ is obtained as 
	\begin{align*}\label{testestimtor22}
		&\reallywidehat{\Delta_{2,2}}=\reallywidehat{J(U_{2,2})}- \reallywidehat{J(X)}\\
		&=\frac{-8}{N}\sum_{i=1}^{N}\left((\log(1-\frac{i}{N+1}))^{2}(1-\frac{i}{N+1})^{2}\left(\frac{2m/N}{X_{i+m:N}-X_{i-m:N}}\right)\right)\\ & \ \ \ \ \ \  \ \ \ \ \ \ \ \ \ \ \ \ \ \ \ \ \ \ \ \ \ \ \ \ + \frac{1}{2N} \sum_{i=1}^{N} \frac{c_i m/N}{X_{i+m:N}-X_{i-m:N}} \\
		&= -\frac{1}{2N} \sum_{i=1}^{N} \left\{32\left(\log(1-\frac{i}{N+1})\right)^2 \left(1-\frac{i}{N+1}\right)^2 - c_i \right\} \frac{m/N}{X_{i+m:N}-X_{i-m:N}}
	\end{align*}

	The following Theorem says $\reallywidehat{\Delta}_{2,2}$ is a consistent estimator of $\Delta_{2,2}$ and proof follows from lines of proof of Theorem 1 of Vasicek (1976) \cite{vasicek}
	\begin{theorem}
		Assume that $X_1,\ X_2,\,..., X_N$ is a random sample of size $N$ taken from a population with pdf $f$ and cdf $F$. Also, let the variance of the random variable be finite. Then $\reallywidehat{\Delta}_{2,2}$ converges in probability to $\Delta_{2,2}$, that is,  ${\reallywidehat{\Delta}}_{2,2}$ is consistent estimator of  ${\Delta}_{2,2}$, as $N \longrightarrow \infty, \ m\longrightarrow \infty \ \text{and} \ \frac{m}{N} \longrightarrow 0.$ 
	\end{theorem}
	
	\indent Note that the test statistics proposed by Park (1999) \cite{park}, Xiong et al. (2020) \cite{xiongetalexp20}, Xiong et al. (2021) \cite{xiongetalsym21}, Jose and Sathar (2022a \cite{josesatharsym22}) and Jose and Sathar ( 2022b \cite{josesathaeexp22} ) are consistent due to the method given in Vasicek (1976) \cite{vasicek}.
	
	\begin{theorem}
		Let $X_1, X_2, ... , X_N$ be a sequence of iid random variables and let $Y_i=aX_i+b, \ a>0,\ b\in \mathbb{R},\ i=1,2,... ,N.$ Denote the estimator for $\Delta_{2,2}$ based on $X_i$ and $Y_i$ by $\reallywidehat{\Delta_{2,2}^X}$ and $\reallywidehat{\Delta_{2,2}^Y}$, respectively. Then 
		\begin{enumerate}[(i)] 
			\item  E($\reallywidehat{\Delta_{2,2}^Y)}=\frac{1}{a}E(\reallywidehat{\Delta_{2,2}^X)}$
			\item  Var($\reallywidehat{\Delta_{2,2}^Y)}=\frac{1}{a^2}Var(\reallywidehat{\Delta_{2,2}^X)}$
			\item  MSE($\reallywidehat{\Delta_{2,2}^Y)}= \frac{1}{a^3}MSE(\reallywidehat{\Delta_{2,2}^X)}$\\
		\end{enumerate}
		where $E(X),\ Var(X)$ and $MSE(X)$ represent expectation, variance and mean square error of random variable $X$, respectively.
	\end{theorem}
	\textbf{Proof: } 
	\begin{align*}
		&\reallywidehat{\Delta^Y_{2,2}} = -\frac{1}{2N} \sum_{i=1}^{N} \left\{32\left(\log(1-\frac{i}{N+1})\right)^2 \left(1-\frac{i}{N+1}\right)^2 - c_i \right\} \frac{m/N}{Y_{i+m:N}-Y_{i-m:N}} \\
		&= -\frac{1}{2N} \sum_{i=1}^{N} \left\{32\left(\log(1-\frac{i}{N+1})\right)^2 \left(1-\frac{i}{N+1}\right)^2 - c_i \right\} \frac{m/N}{aX_{i+m:N}-aX_{i-m:N}}\\
		&= \frac{1}{a} \reallywidehat{\Delta^X_{2,2}}
	\end{align*}
	Thus, proof is completed because of $\reallywidehat{{\Delta}_{2,2}^Y}=a\reallywidehat{{\Delta}_{2,2}^X}$ and  properties of mean, variance and MSE of $X$.
	
	\section{ Critical values }
	The exact critical values of ${\reallywidehat{\Delta}}_{2,2}$ based on 10,000 samples of different sizes generated from an exponential distribution with rate parameter 1 at significance level $\alpha=0.10$, $\alpha=0.05$ and $\alpha=0.01$ are given in Table 1, 2 and 3 respectively. The critical values are obtained for sample sizes $N=5,10,20,30,40,50,100$ with window sizes $m$ ranging from 1 to 50. The next section deals with the simulation study through which the power of the test statistic is evaluated. 
	
	\begin{center}
		{\bf Table 1}. \small{{Critical values of $|{\reallywidehat{\Delta}}_{2,2}|$ statistics at significance level $\alpha$= 0.10}} 
		
		\resizebox{!}{!}{
			\begin{tabular}{ p{1.0cm} p{1.0cm} p{1.0cm} p{1.0cm} p{1.0cm} p{1.0cm} p{1.0cm} p{1.0cm} } 
				\hline
				$m\backslash N$ & 5 & 10 & 20  & 30  & 40 & 50 & 100 \\
				\hline \\
				1  & 0.7174   & 0.4347  & 0.3197 & 0.2808 & 0.2359 & 0.2087 & 0.1503 \\
				2  & 0.5027   & 0.1613  & 0.1355 & 0.1118 & 0.0977 & 0.0879 & 0.0620 \\				
				3  &\         & 0.1411  & 0.0962 & 0.0842 & 0.0761 & 0.0684 & 0.0491   \\
				4  &\         & 0.1659  & 0.0762 & 0.0721 & 0.0671 & 0.0607 & 0.0447 \\
				5  &\         &\        & 0.0660 & 0.0632 & 0.0611 & 0.0563 & 0.0423   \\
				6  &\         &\        & 0.0641 & 0.0553 & 0.0532 & 0.0532 & 0.0395   \\
				7  &\         &\        & 0.0691 & 0.0509 & 0.0487 & 0.0493 & 0.0393   \\
				8  &\         &\        & 0.0777 & 0.0465 & 0.0457 & 0.0458 & 0.0379  \\
				9  &\         &\        & 0.0925 & 0.0453 & 0.0424 & 0.0426 & 0.0368   \\
				10 &\         &\        &\       & 0.0461 & 0.0384 & 0.0397 & 0.0357   \\
				14 &\         &\        &\       & 0.0697 & 0.0380 & 0.0309 & 0.0323  \\
				15 &\         &\        &\       &\       & 0.0396 & 0.0303 & 0.0315   \\
				19 &\         &\        &\       &\       & 0.0570 & 0.0341 & 0.0279   \\
				20 &\         &\        &\       &\       &\       & 0.0364 & 0.0263   \\
				24 &\         &\        &\       &\       &\       & 0.0510 & 0.0228   \\
				30 &\         &\        &\       &\       &\       &\       & 0.0188   \\
				35 &\         &\        &\       &\       &\       &\       & 0.0188  \\
				40 &\         &\        &\       &\       &\       &\       & 0.0219   \\
				45 &\         &\        &\       &\       &\       &\       & 0.0283   \\
				49 &\         &\        &\       &\       &\       &\       & 0.0345   \\
				
				\hline
		\end{tabular}}\label{table1}
	\end{center}

	\begin{center}
		{\bf Table 2}.  \small{{Critical values of $|{\reallywidehat{\Delta}}_{2,2}|$ statistics at significance level $\alpha$= 0.05}}		
		\resizebox{!}{!}{
			\begin{tabular} {p{1.0cm} p{1.0cm} p{1.0cm} p{1.0cm} p{1.0cm} p{1.0cm} p{1.0cm} p{1.0cm}}  
				\hline
				$m\backslash N$ & 5 & 10 & 20  & 30  & 40 & 50 & 100 \\
				\hline \\
				1  & 1.3223   & 0.8269  & 0.5832 & 0.4767 & 0.3779 & 0.3409 & 0.2232 \\
				2  & 0.6977   & 0.2249  & 0.1897 & 0.1552 & 0.1293 & 0.1132 & 0.0791 \\
				3  &\         & 0.1772  & 0.1257 & 0.1095 & 0.0974 & 0.0873 & 0.0600 \\
				4  &\         & 0.2050  & 0.0962 & 0.0915 & 0.0846 & 0.0768 & 0.0561  \\
				5  &\         &\        & 0.0820 & 0.0781 & 0.0767 & 0.0705 & 0.0526   \\
				6  &\         &\        & 0.0770 & 0.0690 & 0.0668 & 0.0666 & 0.0487  \\
				7  &\         &\        & 0.0829 & 0.0617 & 0.0604 & 0.0611 & 0.0478  \\
				8  &\         &\        & 0.0916 & 0.0557 & 0.0561 & 0.0565 & 0.0463   \\
				9  &\         &\        & 0.1075 & 0.0546 & 0.0507 & 0.0519 & 0.0449   \\
				10 &\         &\        &\       & 0.0550 & 0.0470 & 0.0487 & 0.0436   \\
				14 &\         &\        &\       & 0.0809 & 0.0449 & 0.0371 & 0.0389   \\
				15 &\         &\        &\       &        & 0.0461 & 0.0360 & 0.0380   \\
				19 &\         &\        &\       &\       & 0.0655 & 0.0404 & 0.0332   \\
				20 &\         &\        &\       &\       &\       & 0.0427 & 0.0319   \\
				24 &\         &\        &\       &\       &\       & 0.0581 & 0.0272  \\
				30 &\         &\        &\       &\       &\       &\       & 0.0225 \\
				35 &\         &\        &\       &\       &\       &\       & 0.0221 \\
				40 &\         &\        &\       &\       &\       &\       & 0.0258   \\
				45 &\         &\        &\       &\       &\       &\       & 0.0331  \\
				49 &\         &\        &\       &\       &\       &\       & 0.0395   \\
				\hline
		\end{tabular}}
	\end{center}\label{table2}

	\begin{center}
		{\bf Table 3}.  \small{{Critical values of $|{\reallywidehat{\Delta}}_{2,2}|$ statistics at significance level $\alpha$= 0.01}} 
		\resizebox{!}{!}{
			\begin{tabular}{ p{1.0cm} p{1.0cm} p{1.0cm} p{1.0cm} p{1.0cm} p{1.0cm} p{1.0cm} p{1.0cm} } 
				\hline
				$m\backslash N$ & 5 & 10 & 20  & 30  & 40 & 50 & 100 \\
				\hline \\
				1  & 6.0144   & 4.1581  & 2.2252 & 1.8855 & 1.3925 & 1.1696 & 0.6743 \\
				2  & 1.3442   & 0.4419  & 0.3768 & 0.3016 & 0.2610 & 0.2190 & 0.1315 \\
				3  &\         & 0.2843  & 0.2203 & 0.1915 & 0.1666 & 0.1402 & 0.0882 \\
				4  &\         & 0.3054  & 0.1515 & 0.1426 & 0.1304 & 0.1223 & 0.0839 \\
				5  &\         &\        & 0.1204 & 0.1229 & 0.1191 & 0.1053 & 0.0771 \\
				6  &\         &\        & 0.1094 & 0.0994 & 0.1022 & 0.0966 & 0.0693 \\
				7  &\         &\        & 0.1176 & 0.0888 & 0.0897 & 0.0873 & 0.0705   \\
				8  &\         &\        & 0.1317 & 0.0792 & 0.0814 & 0.0823 & 0.0650   \\
				9  &\         &\        & 0.1441 & 0.0727 & 0.0731 & 0.0761 & 0.0638  \\
				10 &\         &\        &\       & 0.0751 & 0.0665 & 0.0685 & 0.0626  \\
				14 &\         &\        &\       & 0.1049 & 0.0590 & 0.0498 & 0.0530   \\
				15 &\         &\        &\       &\       & 0.0617 & 0.0493 & 0.0527   \\
				19 &\         &\        &\       &\       & 0.0839 & 0.0522 & 0.0458   \\
				20 &\         &\        &\       &\       &\       & 0.0572 & 0.0437   \\
				24 &\         &\        &\       &\       &\       & 0.0731 & 0.0375  \\
				30 &\         &\        &\       &\       &\       &\       & 0.0299 \\
				35 &\         &\        &\       &\       &\       &\       & 0.0293  \\
				40 &\         &\        &\       &\       &\       &\       & 0.0339   \\
				45 &\         &\        &\       &\       &\       &\       & 0.0424  \\
				49 &\         &\        &\       &\       &\       &\       & 0.0486   \\
				
				\hline
		\end{tabular}}\label{table3}
	\end{center}

	\section{ Power of test }

	\begin{center}
		{\bf Table 4}.  \small{{Powers of  ${\reallywidehat{\Delta}}_{2,2}$ statistics against alternative $U(0,1)$ at significance level $\alpha= 0.10$}}\\
		\resizebox{!}{!}{
			\begin{tabular}{ p{1.0cm} p{1.0cm} p{1.0cm} p{1.0cm} p{1.0cm} p{1.0cm} p{1.0cm} p{1.0cm} } 
				\hline
				$m\backslash N$ & 5 & 10 & 20  & 30  & 40 & 50 & 100 \\
				\hline \\
				1  & 0.9587  & 0.1988  & 0.1819  & 0.2077 & 0.2476 & 0.2597 & 0.4680 \\
				2  & 1.000   & 0.8896  & 0.5679  & 0.5521 & 0.5909 & 0.6401 & 0.8254 \\
				3  & \       & 0.9932  & 0.8149 & 0.7164 & 0.7293 & 0.7485 & 0.8937 \\
				4  & \       & 1.000  & 0.9269 & 0.8365 & 0.8095 & 0.8080 & 0.9196 \\
				5  & \       & \      & 0.9805 & 0.9089 & 0.8661 & 0.8603 & 0.9313 \\
				6  & \       & \      & 0.9972 & 0.9558 & 0.9199 & 0.8977 & 0.9378 \\
				7  & \       & \      & 0.9998 & 0.9855 & 0.9515 & 0.9290 & 0.9481 \\
				8  & \       & \      & 1.0000 & 0.9961 & 0.9768 & 0.9523 & 0.9609 \\
				9  & \       & \      & 1.0000 & 0.9991 & 0.9882 & 0.9709 & 0.9663 \\
				10 & \       & \      & \      & 0.9997 & 0.9966 & 0.9866 & 0.9705 \\
				14 & \       & \      & \      & 1.0000 & 1.0000 & 0.9998 & 0.9907 \\
				15  & \      & \      & \      & \      & 1.0000 & 0.9999 & 0.9921 \\
				19  & \      & \      & \      & \      & 1.0000 & 1.0000 & 0.9978 \\
				20  & \      & \      & \      & \      & \      & 1.0000 & 0.9993 \\
				24  & \      & \      & \      & \      & \      & 1.0000 & 0.9998 \\
				25  & \      & \      & \      & \      & \      & \ & 1.0000 \\
				30  & \      & \      & \      & \      & \      & \ & 1.0000 \\
				35  & \      & \      & \      & \      & \      & \ & 1.0000 \\
				40  & \      & \      & \      & \      & \      & \ & 1.0000\\
				45  & \      & \      & \      & \      & \      & \ & 1.0000 \\
				49  & \      & \      & \      & \      & \      & \ & 1.0000 \\
				
				\hline
		\end{tabular}}
	\end{center}\label{table4}

	\begin{center}
		\noindent {\bf Table 5}.  \small{{Powers of  ${\reallywidehat{\Delta}}_{2,2}$ statistics against alternative $U(0,1)$ at significance level $\alpha$= 0.05}} \\ 
		\resizebox{!}{!}{
			\begin{tabular}{ p{1.0cm} p{1.0cm} p{1.0cm} p{1.0cm} p{1.0cm} p{1.0cm} p{1.0cm} p{1.0cm} } 
				\hline
				$m\backslash N$ & 5 & 10 & 20  & 30  & 40 & 50 & 100 \\
				\hline \\
				1  & 0.5524  & 0.0629  & 0.0545  & 0.0609 & 0.0742 & 0.0774 & 0.1579 \\
				2  & 1.000   & 0.6353  & 0.3067  & 0.2882 & 0.3485 & 0.4247 & 0.6890 \\
				3  & \       & 0.9849  & 0.6309 & 0.5439 & 0.5678 & 0.5960 & 0.8189 \\
				4  & \       & 0.9999  & 0.8607 & 0.7166 & 0.6899 & 0.6964 & 0.8536  \\
				5  & \       & \       & 0.9648 & 0.8417 & 0.7848 & 0.7785 & 0.8854 \\
				6  & \       & \       & 0.9947 & 0.9226 & 0.8662 & 0.8390 & 0.9036 \\
				7  & \       & \       & 0.9991 & 0.9734 & 0.9145 & 0.8872 & 0.9088 \\
				8  & \       & \       & 1.0000 & 0.9912 & 0.9603 & 0.9226 & 0.9368 \\
				9  & \       & \       & 1.0000 & 0.9982 & 0.9799 & 0.9533 & 0.9374 \\
				10 & \       & \       & \      & 0.9996 & 0.9926 & 0.9791 & 0.9484 \\
				14 & \       & \       & \      & 1.0000 & 1.0000 & 0.9998 & 0.9796 \\
				15  & \      & \       & \      & \      & 1.0000 & 1.0000 & 0.9854 \\
				19  & \      & \       & \      & \      & 1.0000 & 1.0000 & 0.9987 \\
				20  & \      & \      & \      & \      & \      & 1.0000 & 0.9988 \\
				24  & \      & \      & \      & \      & \      & 1.0000 & 1.0000 \\
				25  & \      & \      & \      & \      & \      & \ & 1.0000 \\
				30  & \      & \      & \      & \      & \      & \ & 1.0000 \\
				35  & \      & \      & \      & \      & \      & \ & 1.0000 \\
				40  & \      & \      & \      & \      & \      & \ & 1.0000\\
				45  & \      & \      & \      & \      & \      & \ & 1.0000 \\
				49  & \      & \      & \      & \      & \      & \ & 1.0000 \\
				
				\hline
		\end{tabular}}
	\end{center}\label{table5}

	\begin{center}
		\noindent{\bf Table 6}.  \small{{Powers of  ${\reallywidehat{\Delta}}_{2,2}$ statistics against alternative $U(0,1)$ at significance level $\alpha$= 0.01}} \\
		
		\resizebox{!}{!}{
			\begin{tabular}{ p{1.0cm} p{1.0cm} p{1.0cm} p{1.0cm} p{1.0cm} p{1.0cm} p{1.0cm} p{1.0cm} } 
				\hline
				$m\backslash N$ & 5 & 10 & 20  & 30  & 40 & 50 & 100 \\
				\hline \\
				1  & 0.0719  & 0.0084  & 0.0076  & 0.0115 & 0.0119 & 0.0086 & 0.0108 \\
				2  & 0.9981  & 0.0612  & 0.0200  & 0.0156 & 0.0237 & 0.0412 & 0.2373 \\
				3  & \        & 0.8347  & 0.1888 & 0.1431 & 0.1793 & 0.1556 & 0.5524 \\
				4  & \        & 0.9986  & 0.4906 & 0.2728 & 0.2867 & 0.3594 & 0.6260 \\
				5  & \        & \       & 0.8247 & 0.6058 & 0.4169 & 0.4377 & 0.7117  \\
				6  & \        & \       & 0.9833 & 0.7358 & 0.6173 & 0.5500 & 0.7638 \\
				7  & \        & \       & 0.9979 & 0.9125 & 0.7548 & 0.7145 & 0.7896 \\
				8  & \        & \       & 1.0000 & 0.9774 & 0.8642 & 0.7800 & 0.8301 \\
				9  & \        & \       & 1.0000 & 0.9936 & 0.9299 & 0.8670 & 0.8293 \\
				10 & \       & \        & \      & 0.9998 & 0.9797 & 0.9309 & 0.8693 \\
				14 & \       & \        & \      & 1.0000 & 1.000  & 0.9988 & 0.9358 \\
				15 & \       & \        & \      & \      & 1.0000 & 0.9992 & 0.9615 \\
				19 & \       & \        & \      & \      & 1.0000 & 1.0000 & 0.9919 \\
				20  & \      & \      & \      & \        & \      & 1.0000 & 0.9952 \\
				24  & \      & \      & \      & \        & \      & 1.0000 & 0.9998 \\
				25  & \      & \      & \      & \      & \      & \ & 1.0000 \\
				30  & \      & \      & \      & \      & \      & \ & 1.0000 \\
				35  & \      & \      & \      & \      & \      & \ & 1.0000 \\
				40  & \      & \      & \      & \      & \      & \ & 1.0000\\
				45  & \      & \      & \      & \      & \      & \ & 1.0000 \\
				49  & \      & \      & \      & \      & \      & \ & 1.0000 \\
				
				\hline
		\end{tabular}}
	\end{center} \label{table6}

	\begin{center}
		\noindent{\bf Table 8}.  \small{{Size of ${\reallywidehat{\Delta}}_{2,2}$ for $N=20,50,100$ and significance level $\alpha= 0.05$ for  weibul distribution distribution. }} \\		
		\vspace{0.2cm}
		\resizebox{!}{!}{
			\begin{tabular}{ p{0.5cm} p{0.5cm} p{1.5cm} p{0.5cm} p{0.5cm} p{1.5cm} p{0.5cm} p{0.5cm} p{1.5cm}} 
				\hline
				N   &m       & $w(1,1)$ 	  &N & m   & $w(1,1)$   &N & m   & $w(1,1)$ \\
				\hline 
				\ & 1       & 0.0291          &\ & 1    & 0.0297   &\ & 1       & 0.0296  \\
				\ & 2       & 0.0316          &\ & 2    & 0.0276  &\ & 2        & 0.0342  \\
				\ & 3       & 0.0358          &\ & 3    & 0.0345  &\ & 6        & 0.0371  \\
				\ & 4       & 0.0504          &\ & 5    & 0.0420  &\ & 10       & 0.0364  \\
				\ & 5       & 0.1025          &\ & 8    & 0.0405  &\ & 15       & 0.0450   \\
				20 & 6      & 0.2724          &50 & 10  & 0.0452  &100 & 17     & 0.0501  \\
				\ & 7       & 0.5928          &\ & 15   & 0.1197  &\ & 25       & 0.0512  \\
				\ & 8       & 0.8650          &\ & 20   & 0.6647  &\ & 30       & 0.0731  \\	
				\ & 9       & 0.9915          &\ & 24   & 0.9707  &\ & 40       & 0.4486  \\	
				\ &\        &\		          &\ &\     &\        &\ & \        & \        \\	    
				\hline

		\end{tabular}}
	\end{center}\label{table8}

	\begin{center}
		\noindent{\bf Table 9}.\small{{ Proposed value of window size m for different sample size N }} \\

		\resizebox{!}{!}{
			\begin{tabular}{ p{2.0cm} p{1.0cm}   } 
				\hline
				N  & m   \\
				\hline 
				
				$\leq$10 & 3-5 \\
				10-20    & 6-8 \\	
				21-40     & 11-14  \\	
				41-60     & 15-18 \\	
				60-100    & 20-22  \\
				$\geq$ 100 & 25  \\
				\hline
		\end{tabular}}
	\end{center}\label{table9}

	\section{ Real data application }	
	In this section, we apply our test to detect the suitability of exponentiality on seven real-life data set. A graphical representation of good fit models and the exponential distribution fitted to the data sets 1,2,3 and 4  are presented using histogram and Q–Q plots in Jose and star (2022b) and Q-Q plots for data set 5 and data set 6 are given in Xiong et al. (2020). The exponential distribution is a good fit for dataset 5, the Chen distribution is a good fit for dataset 6 and the uniform distribution is a good fit for dataset 7 (see Xiong et al. (2020)). Every test was conducted at a 5\% nominal level, and 10,000 replications were used for every simulation. Consider the seven datasets given below. 
	

	\textbf{Data set 1}: 74, 57, 48, 29, 502, 12, 70, 21, 29, 386, 59, 27, 153, 26, 326.\\

	Dataset 1 is taken from Proschan (1963) and represents the times between successive failures of air conditioning equipment in a Boeing 720 airplane. The exponential distribution has been used to model this data set (see Shanker et al.(2015), Jose and Sathar (2022b)). For sample size is $N=15$ and window size is $m=3$, the value of test statistics based on dataset 1 is 2.0728 and the corresponding p-value is 0.9992. Our test also detects exponential distribution as a suitable model for this dataset.\\
	
	\textbf{Data set 2}: 12, 17, 7, 13, 5, 2, 12, 2, 6, 4, 5, 14, 6, 2, 4, 18, 4, 19, 5, 14, 20, 8, 11, 26, 1, 3, 10, 18, 6, 10, 23, 7, 20, 4, 7, 6, 12, 10, 20, 3, 12, 3, 18, 18, 14, 14, 8, 6, 22, 11, 8.\\
	
	Data set 2 consists of 51 observations that represent the average maximum temperature (in degrees Celsius) for the 51 major US cities. The National Climatic Data Center (NCDC) of the USA produces the information, which is made available on the website https://www.ncdc.noaa.gov and the Lindley distribution has been used to model this data set (see Jose and Sathar (2022b), Thomas and Jose (2020a)). For sample size is $N=51$ and window size is $m=25$, the value of test statistics based on dataset 2 is 0.5268 and the corresponding p-value is 0.0038. Our test suggests that exponential distribution does not fit well for this dataset.\\
	
	\textbf{Data set 3}: 10.49, 8.8, 12.42, 4.58, 6.85, 4.58, 5., 4.75, 4.75, 12.25, 9.5, 13.54, 10.42, 4.65, 9.88, 6.21, 8.6, 7.06, 7.96, 7.89, 9.7, 13.9, 12.65, 10., 12.65, 12.07, 9.8, 13.54, 9.82, 13.54, 12.42, 12.73, 12.22, 12.25, 12.32, 8.75, 12., 17.5, 11.88, 13.13, 13.56, 15.44, 13.22, 7.28, 11.7, 11.7, 11.6, 10.9, 11.84, 8., 10.2, 5.77, 13.9, 4.58, 12.07, 15.44, 10.2, 11., 8.5, 10.99, 10.39, 9.9, 13.94, 15.21, 13.56, 9., 20.47, 15.22, 11.5, 13.9, 13.22, 10.48, 15.48, 9.8, 12.21, 13.56, 7.04.\\
	
	Data set 3 is taken from Thomas and Jose (2020a). Jose and Sathar (2022b) also used this data set in testing exponentiality. Data set 3 comprises of 77 observations recorded from geoelectrically derived parameters representing	aquifer thickness and The two-parameter Weibull distribution is a good fit for this data set (see Thomas and Jose (2020a), Jose and Sathar (2022b)). For sample size is $N=77$ and window size is $m=30$, the value of test statistics based on dataset 3 is 1.1914 and the corresponding p-value is 0.0051. Our test verifies that exponential distribution does not fit well for this dataset.\\
	
	\textbf{Data set 4}: 0.08, 2.09, 3.48, 4.87, 6.94, 8.66, 13.11, 23.63, 0.2, 2.23, 3.52, 4.98, 6.97, 9.02, 13.29, 0.4, 2.26, 3.57, 5.06, 7.09, 9.22, 13.8, 25.74, 0.5, 2.46, 3.64, 5.09, 7.26, 9.47, 14.24, 25.82, 0.51, 2.54, 3.7, 5.17, 7.28, 9.74, 14.76, 6.31, 0.81, 2.62, 3.82, 5.32, 7.32, 10.06, 14.77, 32.15, 2.64, 3.88, 5.32, 7.39, 10.34, 14.83, 34.26, 0.9, 2.69, 4.18, 5.34, 7.59, 10.66, 15.96, 36.66, 1.05, 2.69, 4.23, 5.41, 7.62, 10.75, 16.62, 43.01, 1.19, 2.75, 4.26, 5.41, 7.63, 17.12, 46.12, 1.26, 2.83, 4.33, 5.49, 7.66, 11.25, 17.14, 79.05, 1.35, 2.87, 5.62, 7.87, 11.64, 17.36, 1.4, 3.02, 4.34, 5.71, 7.93, 11.79, 18.1, 1.46, 4.4, 5.85, 8.26, 11.98, 19.13, 1.76, 3.25, 4.5, 6.25, 8.37, 12.02, 2.02, 3.31, 4.51, 6.54, 8.53, 12.03, 20.28, 2.02, 3.36, 6.76, 12.07, 21.73, 2.07, 3.36, 6.93, 8.65, 12.63, 22.69.\\
	
	Dataset 4 is taken from Linhart and Zucchini (1986). This data set represents the failure times of the air conditioning system of an airplane. The exponential distribution is a good fit for this dataset (see Linhart and Zucchini (1986), Shanker et al.(2015), Jose and Sathar (2022b)). For sample size is $N=128$ and window size is $m=9$, the value of test statistics based on dataset 4 is 255.8024 and the corresponding p-value is 0.382. Our test verifies that exponential distribution is a good fit for this dataset. \\
	
	\textbf{Data set 5}: 5.1, 1.2, 1.3, 0.6, 0.5, 2.4, 0.5, 1.1, 8.0, 0.8, 0.4, 0.6, 0.9, 0.4, 2.0, 0.5, 5.3, 3.2, 2.7, 2.9, 2.5, 2.3, 1.0, 0.2, 0.1, 0.1, 1.8, 0.9, 2.0, 4.0, 6.8, 1.2, 0.4, 0.2.\\
	
	Data set 5 is taken from Bhaumik and Gibbons (2006). This dataset represents the vinyl chloride data obtained from clean-up gradient monitoring wells. This dataset has been fitted very well by exponential distribution (see Xiong et al.(2020), Bhaumik and Gibbons (2006), Shanker et al.(2015) and Marange and Qin (2019)). For sample size is $N=34$ and window size is $m=3$, the value of test statistics based on dataset 5 is 1734.4354 and the corresponding p-value is 0.8335. Our test fails to reject the null hypothesis and therefore, the exponential distribution is a good fit for this dataset.\\

	\textbf{Data set 6}: 0.014, 0.034, 0.059, 0.061, 0.069, 0.080, 0.123, 0.142, 0.165, 0.210, 0.381, 0.464, 0.479, 0.556, 0.574, 0.839, 0.917, 0.969, 0.991, 1.064, 1.088, 1.091, 1.174, 1.270, 1.275, 1.355, 1.397, 1.477, 1.578, 1.649, 1.702, 1.893, 1.932, 2.001, 2.161, 2.292, 2.326, 2.337, 2.628, 2.785, 2.811, 2.886, 2.993, 3.122, 3.248, 3.715, 3.790, 3.857, 3.912, 4.100.\\
	
	Data set 6 is taken from Lawless (2011) and it represents the number of thousands of cycles to failure for electrical appliances in a life test. It contains 50 observations.  For sample size $N=50$ and window size $m=20$, the value of test statistics based on dataset 6 is 5.9071 and the corresponding p-value is 0.0. Our test rejects the null hypothesis even at the 1\% level of significance. This indicates that exponential distribution does not fit this dataset. Chen distribution is a better fit for this dataset than exponential distribution (see Xiong et al.(2020), Yousaf et al. (2019)). \\
	
	\textbf{Data set 7}: 0.0518, 0.0518, 0.1009, 0.1009, 0.1917, 0.1917, 0.1917, 0.2336, 0.2336, 0.2336, 0.2733, 0.2733, 0.3467, 0.3805, 0.3805, 0.4126, 0.4431, 0.4719, 0.4719, 0.4993, 0.6162, 0.6550, 0.6550, 0.7059, 0.7211, 0.7356, 0.7623, 0.7863, 0.8178, 0.8810, 0.9337, 0.9404, 0.9732, 0.9858.\\
	
	Data set 7 is taken from Xiong et al. (2020) and uniform
	distribution is good-fit for it.	For sample size is $N=34$ and window size is $m=5$, the value of test statistics based on dataset 7 is 422.5549 and the corresponding p-value is 0.0001. Our test rejects the null hypothesis even at the 1\% level of significance. This indicates that exponential distribution does not fit this dataset. \\
	
	See Table 11 for the value of test statistics and p-value for different datasets based on the specific window size and sample size of each dataset.\\
	
	\begin{center}
		\noindent{\bf Table 11}.\small{{ Description of models fitted }} 
		
		\resizebox{!}{!}{
			\begin{tabular}{ p{2.0cm} p{1.0cm} p{1.0cm} p{2.0cm} p{2.0cm}  } 
				\hline
				Dataset & N  & m & ${\reallywidehat{\Delta}}_{2,2}$ & p-value  \\
				\hline 
				
				Dataset 1  & 15  & 3  & 2.0728 & 0.9992 \\
				Dataset 2  & 51  & 25 & 0.5268 & 0.0038 \\	
				Dataset 3  & 77  & 30 & 1.1914 & 0.0051 \\
				Dataset 4  & 128 & 9  & 255.8024 & 0.3820 \\
				Dataset 5  & 34  & 3  & 1734.4354 & 0.8335 \\
				Dataset 6  & 50  & 20 & 5.9071 & 0.0000 \\
				Dataset 7  & 34  & 5  & 422.5549 & 0.0001 \\
				\hline
		\end{tabular}}
	\end{center}\label{table11}	
	When testing at a 5\% level of significance, a $p$-value of less than 0.05 indicates that the data does not have exponentiality, whereas a $p$-value of more than 0.05 indicates that the data is exponential distribution as a suitable model. Table 11 indicates that the newly proposed test identifies the exponentiality or non-exponentiality of the distribution of the random sample. The p-values show that datasets 2, 3, 6 and 7 do not have exponentiality in the distribution of the random sample at a 5\% level of significance. Similar to this, a moderate p-value implies that the distributions of datasets 1, 4, and 5 are exponential. We were able to verify that the test statistic correctly identified the exponentiality in the random variable's distribution as a consequence.	
	\section{ Conclusion }
	In the current work, the extropy and an ordered random variable (upper k records) were utilized to characterize the exponential distribution. The test statistic for determining exponentiality was built using this characterisation result for the exponential distribution. . The simulation study is carried out using the Monte Carlo technique to find the value of the estimator, critical value and power of the test. In the majority of cases, it is not the poorest. To demonstrate that the suggested test may be used in practice with confidence, we also included seven examples from real life. \\
	\\
	\\
	\noindent \textbf{ \Large Funding} \\
	\\
	Santosh Kumar Chaudhary would like to thank the Council Of Scientific And Industrial Research (CSIR), Government of India (File Number 09/0081 (14002)/2022-EMR-I) for financial assistance.\\
	\\
	\textbf{ \Large Conflict of interest} \\
	\\
	The authors declare no conflict of interest.\\

	\section*{Appendix}\label{appendix}
	The following steps were used to determine the critical values and compute the power of our proposed test and that of other tests for symmetry at significance level $\alpha=0.10, \ \alpha=0.05, \alpha=0.01:$\\
	(1) we defined a function to calculate the absolute value of ${\reallywidehat{\Delta}}_{2,2}$ .\\
	(2) Generate a sample of size $N$ from the null distribution and compute the test statistics for the sample data;\\
	(3) Repeat Step 2 for 10,000 times and determine the 950th, 975th and 995th quantile respectively of the test statistics as the critical value;\\
	(4) Generate a sample of size $N$ from an alternative distribution and check if the absolute value of the test statistic is greater than the critical value;\\
	(5) Repeat Step 4 for 10,000 times and the percentage of rejection is the power of the test.

	\vspace{.3in}
	
	\noindent
	{\bf  Nitin Gupta} \\
	Department of Mathematics,\\
	Indian Institute of Technology Kharagpur\\
	Kharagpur-721302, INDIA\\
	E-mail: nitin.gupta@maths.iitkgp.ac.in
	
	\vspace{.1in}
	\noindent
	{\bf Santosh Kumar Chaudhary}\\
	Department of Mathematics,\\
	Indian Institute of Technology Kharagpur\\
	Kharagpur-721302, INDIA\\
	E-mail: skchaudhary1994@kgpian.iitkgp.ac.in\\

\end{document}